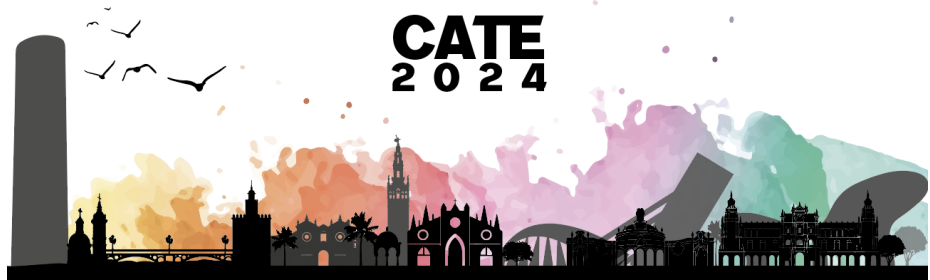

**COMFORT AT THE EXTREMES 2024**
INVESTING IN WELL-BEING IN A CHALLENGING FUTURE

# Motivations and Actions of Human-Building Interactions from Environmental Momentary Assessments

[1*]Pathak, M.P., [2,]Kang S.K., [1]Whittem, V.C., [1]Bassett, K. & [1]Kane, M.B., [3,1]Fannon, D.J.
*lead presenter pathak.m@northeastern.edu
[1] Northeastern University, Civil and Environmental Engineering Department, Boston, MA 02115, United States
[2] Korea Aerospace University, Department of Mechanical and Aircraft System Engineering, Goyang, Gyeonggi-do Province 10540, Republic of Korea
[3] Northeastern University, School of Architecture, Boston, MA 02215, United States

## Abstract

The expansion of renewable electricity generation, growing demands due to electrification, greater prevalence of working from home, and increasing frequency and severity of extreme weather events, will place new demands on the electric supply and distribution grid. Broader adoption of demand response programs (DRPs) for the residential sector may help meet these challenges; however, experience shows that occupant overrides in DRPs compromises their effectiveness. There is a lack of formal understanding of how discomfort, routines, and other motivations affect DRP overrides and other related human building interactions (HBI). This paper reports preliminary findings from a study of 20 households in Colorado and Massachusetts, US over three months. Participants responded to ecological momentary assessments (EMA) triggered by thermostat interactions and at random times throughout the day. EMAs included Likert-scale questions of thermal preference, preference intensity, and changes to 7 different activity types that could affect thermal comfort, and an opened ended question about motivations of such actions. Twelve tags were developed to categorize motivation responses and analyzed statistically to identify associations between motivations, preferences, and HBI actions. Reactions to changes in the thermal environment were the most frequently observed motivation (118 of 220 responses). On the other hand, almost half (47%) responses were at least partially motivated by non-thermal factors, suggesting limited utility for occupant behavior models founded solely on thermal comfort. Changes in activity level and clothing were less likely to be reported when EMAs were triggered by thermostat interactions, while fan interactions were more likely. Windows, shades, and portable heater interactions had no significant dependence on how the EMA was triggered. These results suggest that better understanding of motivations for HBI may improve effectiveness of demand response programs.

**Introduction**
Many forces demand modernization of electricity grid infrastructure and control to improve sustainability and increase resilience. Pressures include fluctuations in the supply of electricity as the share of renewable energy generation increases (Denholm et al., 2014; Langevin et al., 2021), increasing electrification of building and transportation sectors (Nadel, 2019), and the increased incidence of extreme weather events (Gao et al., 2018; Zamuda et al., 2018). Demand side management, a collective term for programs that flexibly control demand based on electricity availability, is widely regarded as an important tool to balance the sustainability and resilience of the electric grid (Chen et al., 2018). In addition, demand side management assists utilities to maintain the power supply during extreme weather events (Pratt & Ericsen 2020)

Demand Response (DR) is a particular kind of demand side management that incrementally changes thermostat set points or uses smart thermostats to shift load timing; for example, in the residential sector, by reducing peak cooling demand in the summer. The DR provider (usually the electric utility) may *directly* adjust the thermostat setpoint or turn off the HVAC system for a fixed period without customer interaction (called direct, active or dispatchable control), or the provider may communicate through calls or SMS to *encourage* customers to reduce consumption (called indirect load control). In both cases, occupants may override thermostat setpoints or power switches but forfeit DR incentives when doing so (Tomat et al., 2022).

However, prior research indicates numerous challenges for DR programs. First, up to 30% of households participating in direct load control programs override the thermostat setpoint during DR events (Navigant, 2017). Secondly, during indirect load control events occupant behaviors are unreliable and hard to predict (Pratt & Erickson, 2020). Failure of DR programs to reach peak demand reduction goals decreases electricity grid reliability and increases fossil fuel emissions prompting large financial penalties (Gilbraith & Powers, 2013). In addition, as a result of DR failures utilities can face financial distress and even bankruptcy from purchasing electricity on the wholesale spot market during extreme weather events (Popik & Humphreys, 2021).

Thermal comfort has been shown to be a leading cause of setpoint overrides during DR events (Aghniaey et al., 2018; da Fonseca et al., 2021), prompting the application of numeric thermal comfort models to improve DR programs. However, other motivations, including control over the timing of home activities, control over expenses, and general personal autonomy (Sarran et al., 2021) also drive thermostat setpoint overrides.

This paper describes research to address this limitation. Ecological Momentary Assessment (EMA) surveys were used to investigate relationships between Human Building Interaction (HBI) events and self-reported motivations and provide some preliminary results. The EMA surveys were collected as part of the Whole Energy Homes project from volunteers in their own homes. The project monitored 20 homes in 2 US climates – Massachusetts and Colorado and all adults in each home were included in the EMA survey prompts, sent to their mobile telephones.

*Ecological Momentary Assessment (EMA) Background*
EMAs have been used for many years in psychology, medicine and sociology to obtain real-time assessment of participants' behavior and experiences in their natural environments (Shiffman et al., 2008). Ecological refers to collecting data under real life conditions, rather than mediated by laboratory conditions, and momentary refers to frequent, brief methods, concurrent with the behavior under examination. EMAs are designed to overcome limitations in human autobiographical recall. Research has shown that autobiographical memory is subject to systematic biases, related to humans creating recall from fragmentary memory. Thus, experiences which are more unique or intense are more likely to be retained. (Redelmeier et al., 2003) Further, recall is influenced by beliefs. People unconsciously re-organize their 'memories' to fit a coherent script or theory or to reconcile them with later events (Ross, 1989). Thus, the closer in time an EMA survey is to an event, the more accurate the participants' recall will be. (Shiffman et al., 2008).

While EMA has been used to study occupant thermal comfort (Duarte Roa et al., 2020; Huang et al., 2015) including longitudinally (R. de Dear et al., 2018; Langevin et al., 2015; Langevin, 2019), there are a limited number of studies where EMA has been used to examine energy consumption behavior (Nambiar et al., 2024) .

*Aims*
Residential energy use accounts for a large proportion of US national energy (LLNL, 2024) consumption. Mismatches between electricity supply and demand can be partly managed with Demand Response programs. Demand Response programs used to reduce power consumption suffer from a high rate of householder thermostat overrides. Better understanding Human Building Interactions related to energy consumption in the residential context can improve Demand Response programs and prediction algorithms. Therefore, this study applies the EMA method adapted from other disciplines to examine the detailed motivations and particular actions of participants linked to HBI events, as a preliminary step towards future work developing generalized models of occupant behavior representative of the whole population.

**Method**

*Whole Energy Home Data Collection*
This research is part of the 5-year Whole Energy Homes (WEH) project, funded by the US Department of Energy (Award Number DE-EE0009154). Data collection for the full study has been described in detail in (Casavant et al., 2022). EMA micro-surveys used the Home Assistant application on participants' mobile phones. The WEH study is particularly interested in links between granular qualitative and quantitative data on individual occupants' presence, perception, mental state, and behavior.

*EMA Question Development and Rationale*
To understand the complex relationship between occupant behavior and manual thermostat setpoint changes, the study collected data about thermally influential behaviors and motivations. It aimed to capture *all* behaviors that affect an occupant's heat balance and perception, not only those intended to improve the thermal comfort. Some behaviors affect occupants' thermal experience as a by-product of satisfying a non-thermal need, for example, closing the drapes for privacy on a winter night may also reduce the heat loss through the windows.

Unlike past studies (R. de Dear et al., 2018; Duarte Roa et al., 2020; Sun et al., 2018) which asked about current state, this study asked about state *transitions* within the past hour, for example, "Have you adjusted the position of a window in the last hour" rather than "Is there an open window near you?". Asking about recent changes and actions provided the opportunity to collect data regarding the *motivation* for the change while still supporting accurate recall. The fixed window of one hour aligns with the period an EMA survey is open and gives an approximate time of when the change was made, enabling comparison to sensor and thermostat data.

In keeping with the goal of a very short survey (<10 seconds to complete), the total number of questions is small. Because many survey prompts were triggered by thermostat interactions, we captured both thermal preference and preference intensity for temperature change. The team selected questions about non-thermostat actions in the previous hour—shown in Figure 1—to capture other factors affecting thermal comfort, for example, clothing and activity levels per ASHRAE (R. J. de Dear & Brager, 2002), use of electrical appliances or other human building interactions. Most important, an open-ended prompt invited participants to freely describe the motivation for the action/s in a one-line response.

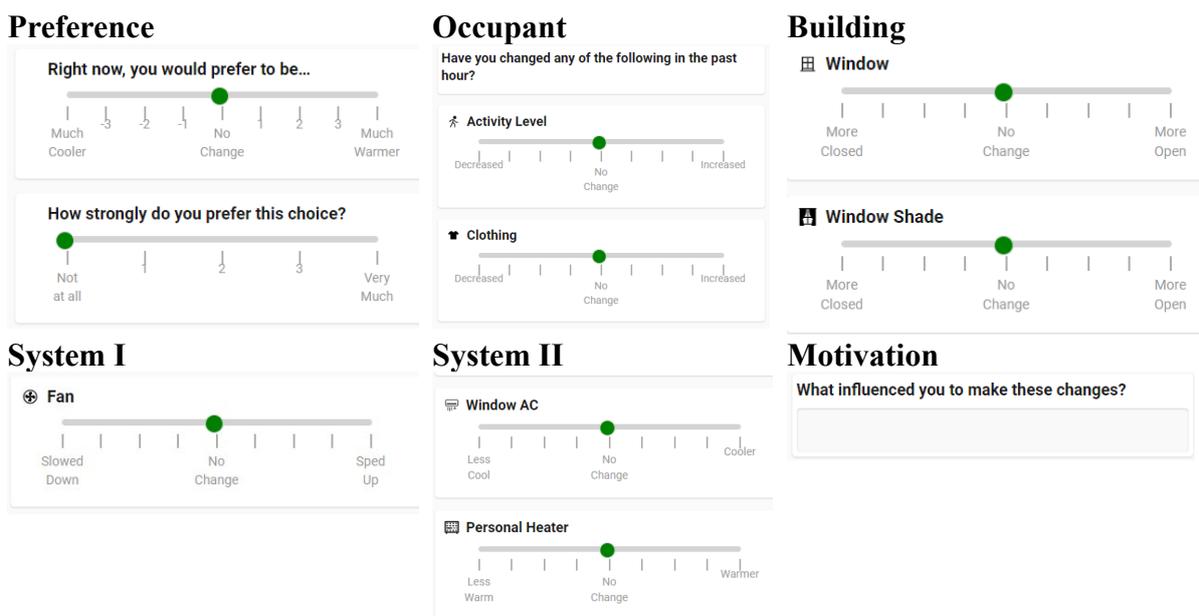

Figure 1: EMA questions asking about occupant preferences, and actions taken during the last hour.

### Collecting EMA Responses

EMA data is collected in three situations 1) push notifications triggered by thermostat adjustments, 2) push notifications generated on a random time basis during the hours spent at home and 3) participants opt to complete a survey whenever they wished. The random push notifications provide control data for times when participants were comfortable, or at least not sufficiently *uncomfortable* to change the thermostat. EMA survey time limits include: no surveys during sleeping hours (between 10pm and 7am), no more than 8 surveys per week, and no surveys within 15 minutes of arriving home, to allow participants to acclimate to indoor temperatures. This paper includes 528 EMA surveys, taken over 75 days, 2023-09-19 to 2023-11-02.

### Analysis - Development of Motivation Tags for Qualitative Responses

Of the 528 EMA responses in the sample, 220 included motivations. All motivation responses were analyzed to develop a limited set of motivation tags for use in statistical analysis. To reduce the opportunity for bias, researchers tagging the qualitative motivation answers were not provided with the rest of the EMA survey data. The Jaccard index was calculated to quantify differences in interpretation between taggers, the result of about 0.8 indicates a roughly 80% similarity between taggers. Cross-checking the remaining tags by the other tagger helped increase the similarity further. Table 1 lists the final motivation tags developed using this method.

Table 12 - Codebook of motivation tags, their descriptions, and example texts.

| Tag | Description | Examples |
|---|---|---|
| thermal_reactive | Behaviors in response to personal thermal discomfort or are intended to improve personal thermal comfort immediately. This may be regarding temperature and/or humidity. This tag is only for the thermal comfort motivations of the person answering the EMA. | *"Too cold"* <br> *"Too hot in here"* |
| thermal_proactive | Behaviors intended to prepare respondents or their home for thermal conditions that will occur in the future. | *"The sun is moving to the west side of the house. Closing the blinds keeps the cooler temps indoors"* |
| routine | Behaviors that imply some regularity in time (time of day, day of week, time of year, etc.). | *"Waking up"* <br> *"Closing shades because of nighttime"* |
| outdoor_response | Behaviors explicitly in response to the outdoor conditions. These conditions may be thermal, visual, etc. | *"The sun is setting and it is getting dark and cold. It was in the 80's today and it's too hot"* |
| financial | Behaviors in response to financial considerations. | *"Cost of electricity"* |
| lighting | Behaviors to change the illumination of the indoor environment. | *"More light as the sun goes down"* |
| transition_indoors | Behaviors associated with occupants transitioning from outdoors to indoors. | *"Came in hot from a walk"* <br> *"Came home from morning drop off walk"* |
| practical_behavior | Behaviors aimed at achieving a specific objective outside of the categories listed here, but which may have ancillary thermal consequences. | *"Wanted to hear a Zoom call"* <br> *"Stretching and packing for a biz trip"* <br> *"Let in more light"* |
| social | Behaviors by the participant answering the EMA, but motivated by another person. The reason that the other person wants the change does not matter. This means that the other person may be motivated by thermal factors, finances, etc. but they will still be tagged with this tag. | *"Wife"* <br> *"Others in the household would complain"* |
| air_movement | Behaviors intended to affect airflow or air quality. | *"I like to open the windows in the am to get fresh air in my house."* |
| air_movement | Behaviors affecting airflow but where that is not the main goal. | *"Room was a little warm without the fan so I turned it on."* |
| convenience | Behaviors motivated or demotivated by how easy or convenient the behavior is. | *"Too busy"* |
| unsure | This tag is for all behavior motivations that are unclear either because they do not seem to fit into any of the above tags, or because they are just repeating the behavior without context as to why. | *"Windows opened"* <br> *"Didn't set it right"* |

## Results and Discussion

*Analysis of Occupant Motivations: Thermal responses and their contextual influences*
Following the qualitative tagging of the "raw text" motivations, as outlined in the methodology section, a total of 220 EMAs. As illustrated in Table 1, each motivation response was tagged with one or more motivation labels, reflecting the complexity of occupant behaviors. The distribution of tags highlights that "thermal reactive" motivations—reported in 118 instances—dominate the dataset. This suggests that thermal discomfort, accumulated over time, is a primary driver behind occupant actions. The second most frequent motivation, "outdoor response," with 36 occurrences, underscores the substantial impact of external environmental conditions on occupant behavior, further emphasizing the dynamic interplay between indoor and outdoor environments in detached, single-family homes.

Motivations driven by "social" factors (19 occurrences) point to the influence of interpersonal interactions and communal dynamics in decision-making, while the presence of "routine" (18) and "practical behavior" (17) motivations reflects the structured, habitual nature of some thermally affective occupant activities, indicating predictable actions in response to environmental cues or personal habits. In the subsequent section, we delve deeper into whether the observed routine behavior intersects with thermostat interactions, an important aspect for understanding occupant-driven control of building systems.

Motivations such as "air movement," "thermal proactive," and "convenience," occurred less frequently, but do indicate the breadth of factors influencing occupant behavior beyond thermal discomfort. The single mention of "financial" concerns, suggests that economic factors rarely directly motivate on a specific action, at least in these data. The presence of "unsure" responses (14 occurrences) where researchers could not to definitively attribute a clear motivation for the reported actions, further illustrating the complexity of human decision-making in environmental contexts.

The histogram in Figure 2 shows that over 220 motivation responses correspond to 259 tagged occurrences, indicating that many open-ended responses were assigned multiple motivations. This multi-dimensionality of motivations reveals that occupant actions, particularly in human-building interactions, are shaped by a broader spectrum of influences than thermal comfort alone. Of the 220 EMAs, only 118 (~54%) motivations were identified as "thermal reactive," and 16 (~7%) as "routine," suggesting that occupant-centric control systems should consider factors beyond thermal discomfort or routine behaviors when designing interfaces and controls for residential settings. This distribution lays the foundation for the analysis of how these motivations correspond to different occupant actions, explored further in the following sections.

*Motivation Distribution in Relation to Actions and EMA Triggers*

To understand the relationship of actions reported by the occupant in the EMAs along with their reported motivations and what prompted them to fill in the EMAs in the first place we performed a two-part analysis, first the relationship between Actions and EMA Types and then

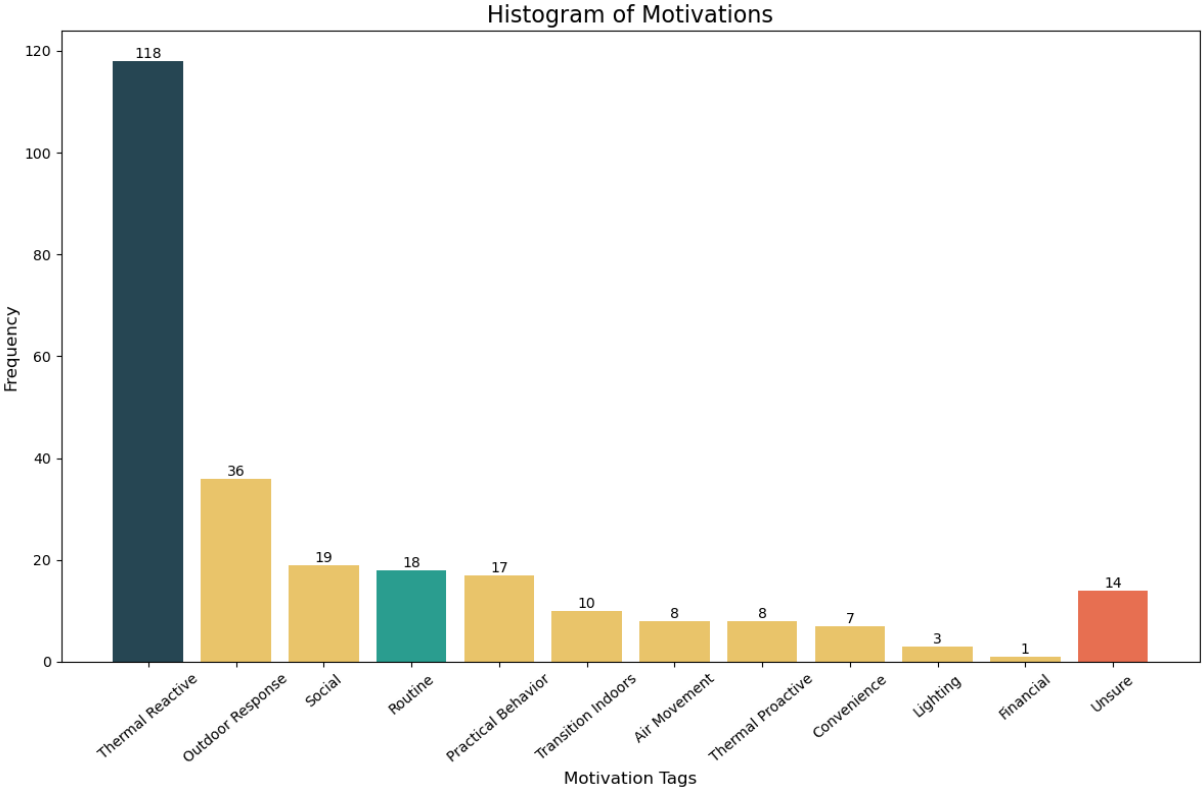

Figure 2: Histogram of Motivation Tags. This figure displays the distribution of various motivation tags assigned to occupant responses across EMAs. While thermally reactive motivations are most frequent (118 of 220 EMA or 54%) nearly as many have the diverse non-thermal motivations listed in table 1 above.

proceeded to analyze the Motivations Categorizations in relation to the actions taken and EMA types:

*Analysis of Actions Taken Based on EMA Type*
The analysis in Figure 3 includes occupant behaviors such as adjusting clothing, changing activity levels, and manipulating environmental controls (e.g., window, shade operation, fan and other secondary HVAC system usage). Based on the analysis of EMA types and reported occupant actions, fewer actions were observed in thermostat-triggered EMAs compared to random EMAs, except for fan interactions. This suggests overall, occupants made fewer changes a priori thermostat interactions, regardless of whether it was an occupant's personal state change such as activity and clothing levels or interaction with building and system components. Conversely, randomly triggered EMAs include a broad range of occupant actions reported, indicating more than just thermal corrective actions. We used chi-square tests to evaluate whether any specific occupant action (e.g., opening or closing a window, indicated by $w_i \neq 0$) was independent of the sampling method (i.e. whether the EMA was triggered by a thermostat interaction or sent randomly). In cases where null hypothesis of this test is rejected (i.e., $p < 0.05$) we can infer that the occupant is statistically more or less likely to perform the action in the hour preceding a thermostat interaction compared to a randomly sampled time. Based on this, three activities – activity level change (p-value = 5.21e-05), clothing changes (p-value = 0.00), and fan interaction (p-value = 2.50e-09) are statistically more likely to be reported when a thermostat interaction driven EMA is reported. The secondary heating system actions are only observed when EMA are randomly sampled, strongly suggesting negative association with thermostat interactions.

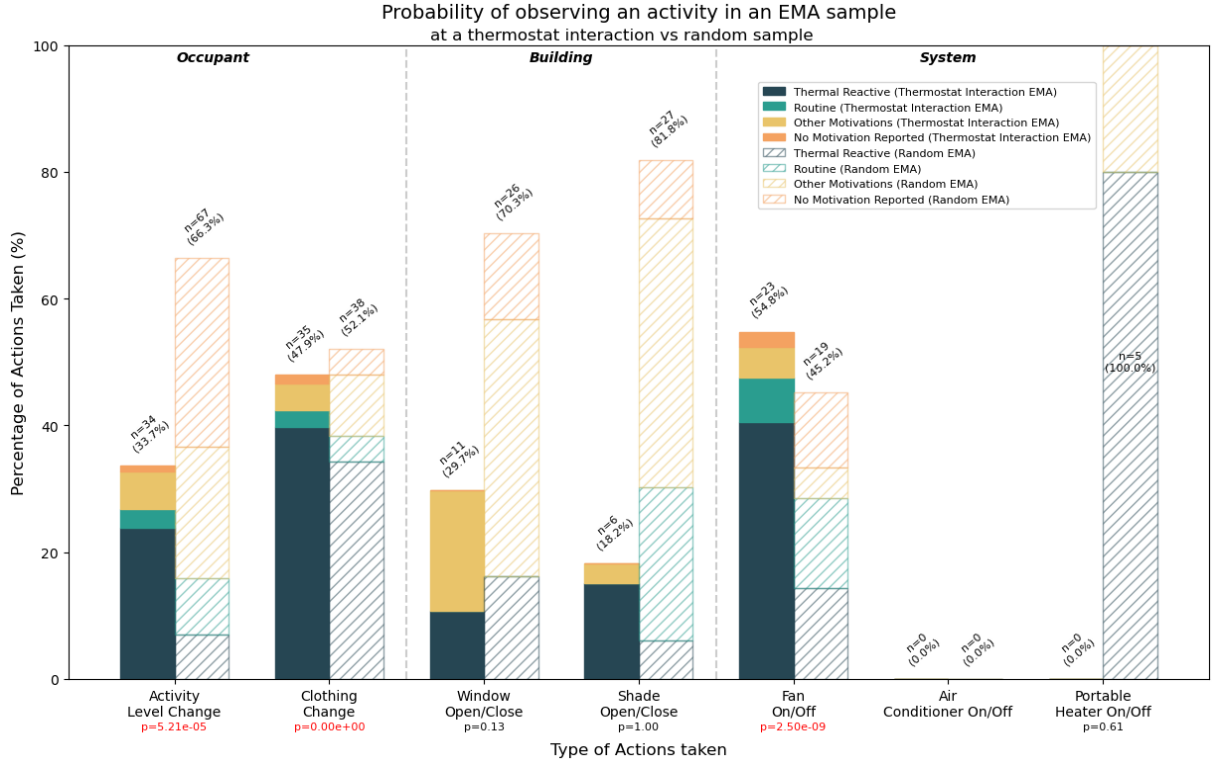

Figure 3: Stacked bar plot displaying the distribution of EMAs based on motivation, divided by the type of EMA trigger and segmented by reported action. For each action, the left bar represents the percentage of EMAs triggered by thermostat interactions, while the right bar represents the percentage triggered randomly. Both bars sum to 100%, encompassing all EMAs that reported actions. Each bar is segmented reflecting the proportion of EMAs linked to distinct motivation categories based on the corresponding tags: 'Thermal Reactive', 'Routine'. 'Other Motivation' represents EMAs with motivations outside the two primary categories, and 'No Motivation Reported' captures instances where occupants did not provide a motivation response.

We further explored the relationship between occupant motivations (e.g., thermal reactive, routine, and other motivations) and the type of EMA (thermostat-triggered or random):

***Thermal Reactive Motivations:***
*Thermostat-triggered EMAs*: Among the EMA's prompted by a thermostat-change, participants reporting a thermally-reactive motivation are more likely to have taken other thermal actions in the past hour than participants whose EMAs after thermostat changes report non-thermally reactive or no motivation. Activity-levels (24 occurrences), Clothing changes (29 occurrences) and fan usage (17 occurrences) were particularly prominent, suggesting that thermal discomfort is a primary driver of actions taken associated with thermostat interactions.

*Random EMAs*: Even in random EMAs, occupants reported some thermal reactive motivations, but the actions taken were less frequent. For example, Activity Level Change (7 occurrences), Clothing changes (25 occurrences) and fan usage (6 occurrences) were lower compared to thermostat triggered EMAs, suggesting that thermal discomfort is less frequently a motivation during random moments.

***Routine Motivations:***
*Thermostat-triggered EMAs*: Routine motivations were less commonly associated with thermostat-triggered EMAs, with relatively fewer instances of personal thermal actions like clothing change (2 occurrences) and activity level changes (3 occurrences). This suggests that routine motivations are not as strongly associated with thermal discomfort as thermal reactive motivations.

*Random EMAs*: In randomly prompted EMAs, routine behaviors were more frequent, indicating that occupants are more likely to engage in daily routines of all actions independently of thermal discomfort. For example, 9 instances of activity level changes were reported in random EMAs driven by routine motivations.

***Other Motivations:***

*Random EMAs*: Many motivations were observed for other various actions during random EMAs, particularly for activity level changes (51 occurrences). This suggests that random EMAs capture a broad range of occupant behaviors reflecting the variety of factors influencing occupant actions that may not relate directly to thermal discomfort or routine.

*Thermostat-triggered EMAs*: In contrast, thermostat-triggered EMAs had a much lower occurrence of actions under the "other motivations" category, highlighting the focus on thermal corrective actions in these moments.

*Secondary heating equipment use*: Heater use is negatively associated with thermostat interactions indicating that participants pursue one or the other activity.

***Implications***
The findings from this study demonstrate critical implications for advancing demand flexibility (DF) research in residential buildings. Our results reveal that current standard practice models for demand flexibility are constrained by their narrow focus on thermal dynamics and comfort metrics alone. These models overlook essential behavioral and contextual variables that significantly shape how occupants respond to DF events. There is a pressing need to develop more comprehensive causal models that capture non-thermal reactive inputs, as our findings show that occupants' adaptive behaviors are often driven by priorities beyond thermal comfort considerations. The significant variation in occupants' responses to temperature adjustments, influenced by factors extending beyond thermal comfort, challenges the traditional paradigm of modeling buildings as purely responsive thermodynamic systems.

Furthermore, this study highlights a critical gap in current building performance modeling practices. While existing models excel at capturing thermal comfort metrics, they often fail to account for the dynamic and adaptive nature of occupant behavior during demand flexibility events. This limitation stems from the traditional focus on steady-state comfort conditions rather than the transient states that characterize DF interventions. The development of more sophisticated building performance models that can capture both thermal reactivity and occupant adaptability is essential for accurately predicting and optimizing demand flexibility outcomes.

These findings point to a need for a paradigm shift in how we approach demand flexibility research, moving from purely physics-based models to integrated frameworks that can capture the complex interplay between building systems, occupant behavior, and environmental conditions.

*Limitations*
The study participants were self-selected in two locations, rather than a random sample, consequently we cannot generalize from these results to the broader population. Five homes in Florida have since joined the study, which will provide data from an additional climate zone. Furthermore, although the number of EMA responses is sufficient for statistical analysis and chi-squared tests, the small proportion of responses relating to some motivation tags limit the conclusions that can be drawn. Aggregating low-frequency motivations as one of the four groups shown in Figure 3, rather than analyzing them individually helps mitigate this limitation.

Other influences—such as financial motivations, environmental attitudes, and social factors-while not examined in this work, were explored in participant interviews which will contribute to future analysis of EMA surveys. The collection period for these EMA surveys largely fell during shoulder season in MA and CO. Future extensions of the study will analyze the motivations, HBI actions and particularly thermostat interactions with a larger sample size over a longer period.

**Conclusion**
This study provides novel insights into human-building interactions based on occupant responses collected through Ecological Momentary Assessments (EMAs) over the period of ~75 days. Findings indicate that thermal discomfort is the most frequent driver of actions related to clothing, activity level, and fan usage with just over half of responses. However, randomly triggered EMAs revealed as many (46%) responses pointing to a broad range of occupant behaviors highlighting the complexity of non-thermal motivations such as routine and social factors. Better predicting human responses to Demand Response events requires better characterization of non-thermal factors, including as family structure, dwelling hours and household routines related to thermostat behavior. In support of that aim, the WEH homes project is conducting a second phase of this study with the same participants in which we control the thermostat to simulate Demand Response events while collecting associated EMA surveys. Improved predictions would enhance demand flexibility programs, improve building performance, and maintain occupant comfort in residential settings especially when homes participate as a demand-response resource in year-round grid interactions.

**Acknowledgements**
This research was funded by the US Department of Energy DE-EE0009154 and US NSF CAREER award #2047317. We gratefully acknowledge our research participants. All participants are adults, who provided informed consent for themselves, consistent with relevant guidelines and regulations for the protection of human subjects per research protocol (IRB #21-07-01) approved by the Northeastern University Institutional Review Board.

**References**

Aghniaey, S., Lawrence, T. M., Mohammadpour, J., Song, W., Watson, R. T., & Boudreau, M. C. (2018). Optimizing thermal comfort considerations with electrical demand response program implementation. *Building Services Engineering Research and Tecnology*, *39*(2), 219–231. https://doi.org/10.1177/0143624417752645

Casavant, E., Pathak, M., Fannon, D., Sharma, K., Pavel, M., & Kane, M. (2022). A Novel Methodology for Longitudinal Studies of Home Thermal Comfort Perception and Behavior. *2022 Summer Study Proceedings*. 2022 ACEEE Summer Study on Energy Efficiency in Buildings. https://aceee2022.conferencespot.org/event-data/pdf/catalyst_activity_32550/catalyst_activity_paper_20220810191618100_eb740d6f_f63a_4f7b_9895_407c6e6dbb70

Chen, Y., Xu, P., Gu, J., Schmidt, F., & Li, W. (2018). Measures to improve energy demand flexibility in buildings for demand response (DR): A review. *Energy and Buildings*, *177*, 125–139. https://doi.org/10.1016/j.enbuild.2018.08.003

da Fonseca, A. L. A., Chvatal, K. M. S., & Fernandes, R. A. S. (2021). Thermal comfort maintenance in demand response programs: A critical review. *Renewable and Sustainable Energy Reviews*, *141*, 110847. https://doi.org/10.1016/j.rser.2021.110847

de Dear, R. J., & Brager, G. S. (2002). Thermal comfort in naturally ventilated buildings: Revisions to ASHRAE Standard 55. *Energy and Buildings*, *34*(6), 549–561. https://doi.org/10.1016/S0378-7788(02)00005-1

de Dear, R., Kim, J., & Parkinson, T. (2018). Residential adaptive comfort in a humid subtropical climate—Sydney Australia. *Energy and Buildings*, *158*, 1296–1305. https://doi.org/10.1016/j.enbuild.2017.11.028

Denholm, P., Margolis, R., Palmintier, B., Barrows, C., Ibanez, E., Bird, L., & Zuboy, J. (2014). *Methods for Analyzing the Benefits and Costs of Distributed Photovoltaic Generation to the U.S. Electric Utility System* (NREL/TP-6A20-62447). National Renewable Energy Lab. (NREL), Golden, CO (United States). https://doi.org/10.2172/1159357

Duarte Roa, C., Schiavon, S., & Parkinson, T. (2020). Targeted occupant surveys: A novel method to effectively relate occupant feedback with environmental conditions. *Building and Environment*, *184*, 107129. https://doi.org/10.1016/j.buildenv.2020.107129

Gao, X., Schlosser, C. A., & Morgan, E. R. (2018). Potential impacts of climate warming and increased summer heat stress on the electric grid: A case study for a large power transformer (LPT) in the Northeast United States. *Climatic Change*, *147*(1), 107–118. https://doi.org/10.1007/s10584-017-2114-x

Gilbraith, N., & Powers, S. E. (2013). Residential demand response reduces air pollutant emissions on peak electricity demand days in New York City. *Energy Policy*, *59*, 459–469. https://doi.org/10.1016/j.enpol.2013.03.056

Huang, C.-C. (Jeff), Yang, R., & Newman, M. W. (2015). The potential and challenges of inferring thermal comfort at home using commodity sensors. *Proceedings of the 2015 ACM International Joint Conference on Pervasive and Ubiquitous Computing*, 1089–1100. https://doi.org/10.1145/2750858.2805831

Kihlstrom, J. F., Eich, E., Sandbrand, D., & Tobias, B. A. (1999). Emotion and memory: Implications for self-report. In *The science of self-report* (pp. 93–112). Psychology Press.

Langevin, J. (2019). Longitudinal dataset of human-building interactions in U.S. offices. *Scientific Data*, *6*(1), 288. https://doi.org/10.1038/s41597-019-0273-5

Langevin, J., Gurian, P. L., & Wen, J. (2015). Tracking the human-building interaction: A longitudinal field study of occupant behavior in air-conditioned offices. *Journal of Environmental Psychology*, *42*, 94–115. https://doi.org/10.1016/j.jenvp.2015.01.007

Langevin, J., Harris, C. B., Satre-Meloy, A., Chandra-Putra, H., Speake, A., Present, E., Adhikari, R., Wilson, E. J. H., & Satchwell, A. J. (2021). US building energy efficiency and flexibility as an electric grid resource. *Joule*, *5*(8), 2102–2128. https://doi.org/10.1016/j.joule.2021.06.002

Nadel, S. (2019). Electrification in the Transportation, Buildings, and Industrial Sectors: A Review of Opportunities, Barriers, and Policies. *Current Sustainable/Renewable Energy Reports*, *6*(4), 158–168. https://doi.org/10.1007/s40518-019-00138-z

Nambiar, C., Rosenberg, S., Peffer, T., Schiavon, S., Brager, G., Zhang, H., & Rees, A. (2024). *Enhancing Participation in Residential Demand Response: Insights from Case Studies Conducted in Alaska and California*.

Navigant. (2017). *National Grid Smart Energy Solutions Pilot—Final Evaluation Report*. Massachusetts Electric Company and Nantucket Electric Company d/b/a National Grid. https://fileservice.eea.comacloud.net/FileService.Api/file/FileRoom/9179984

Popik, T., & Humphreys, R. (2021). The 2021 Texas Blackouts: Causes, Consequences, and Cures. *Journal of Critical Infrastructure Policy*, *2*(1), 47–73. https://doi.org/10.18278/jcip.2.1.6

Pratt, B. W., & Erickson, J. D. (2020). Defeat the Peak: Behavioral insights for electricity demand response program design. *Energy Research & Social Science*, *61*, 101352. https://doi.org/10.1016/j.erss.2019.101352

Redelmeier, D. A., Katz, J., & Kahneman, D. (2003). Memories of colonoscopy: A randomized trial. *Pain*, *104*(1–2), 187–194.

Ross, M. (1989). Relation of implicit theories to the construction of personal histories. *Psychological Review*, *96*(2), 341.

Sarran, L., Gunay, H. B., O'Brien, W., Hviid, C. A., & Rode, C. (2021). A data-driven study of thermostat overrides during demand response events. *Energy Policy*, *153*, 112290. https://doi.org/10.1016/j.enpol.2021.112290

Shiffman, S., Stone, A. A., & Hufford, M. R. (2008). Ecological Momentary Assessment. *Annual Review of Clinical Psychology*, *4*(1), 1–32. https://doi.org/10.1146/annurev.clinpsy.3.022806.091415

Sun, C., Zhang, R., Sharples, S., Han, Y., & Zhang, H. (2018). A longitudinal study of summertime occupant behaviour and thermal comfort in office buildings in northern China. *Building and Environment*, *143*, 404–420. https://doi.org/10.1016/j.buildenv.2018.07.004

Tomat, V., Vellei, M., Ramallo-González, A. P., González-Vidal, A., Le Dréau, J., & Skarmeta-Gómez, A. (2022). Understanding patterns of thermostat overrides after demand response events. *Energy and Buildings*, *271*, 112312. https://doi.org/10.1016/j.enbuild.2022.112312

Zamuda, C. D., Bilello, D. E., Conzelmann, G., Mecray, E., Satsangi, A., Tidwell, V., & Walker, B. J. (2018). Chapter 4: Energy Supply, Delivery, and Demand. In S. C. Pryor (Ed.), *Fourth National Climate Assessment, Volume II: Impacts, Risks, and Adaptation in the United States* (pp. 1–470). U.S. Global Change Research Program, Washington, DC. https://nca2018.globalchange.gov/chapter/4/